\documentclass[twocolumn,preprintnumbers,prl,floatfix,superscriptaddress,amsmath,amssymb]{revtex4}

\usepackage{graphicx}
\usepackage{dcolumn}
\usepackage{bm}
\usepackage{dsfont}
\usepackage{color}

\begin{document}

\title{Synchronization invariance under network structural transformations}

\author{Llu\'is Arola-Fern\'andez}
\affiliation{Departament d'Enginyeria Inform\`atica i Matem\`atiques, Universitat Rovira i Virgili, 43007 Tarragona, Spain}
\author{Albert D\'iaz-Guilera} 
\affiliation{Departament de F\'isica Fonamental, Universitat de Barcelona, Mart\'i i Franqu\`es 1, 08028 Barcelona, Spain}
\affiliation{Universitat de Barcelona Institute for Complex Systems (UBICS), Barcelona, Spain}
\author{Alex Arenas}
\affiliation{Departament d'Enginyeria Inform\`atica i Matem\`atiques, Universitat Rovira i Virgili, 43007 Tarragona, Spain}

\date{\today}

\begin{abstract}
Synchronization processes are ubiquitous despite the many connectivity patterns that complex systems can show. Usually, the emergence of synchrony is a macroscopic observable, however, the microscopic details of the system, as e.g. the underlying network of interactions, is many times partially or totally unknown.
We already know that different interaction structures can give rise to a common functionality, understood as a common macroscopic observable. Building upon this fact, here we propose network transformations that keep the collective behavior of a large system of Kuramoto oscillators functionally invariant. We derive a method based on information theory principles, that allows us to adjust the weights of the structural interactions to map random homogeneous -in degree- networks into random heterogeneous networks and vice-versa, keeping synchronization values invariant. The results of the proposed transformations reveal an interesting principle; heterogeneous networks can be mapped to homogeneous ones with local information, but the reverse process needs to exploit higher-order information. The formalism provides new analytical insight to tackle real complex scenarios when dealing with uncertainty in the measurements of the underlying connectivity structure.
\end{abstract}

\maketitle

The study of dynamical processes running on top of complex networks has become a central issue in many research fields, ranging from the microscopic realm of genes and neurones to the large realm of technological and social systems \cite{strogatz01,boccaletti06,doro08,arenas08,newman10,pastor15,dedomenico16a}. The interplay between topology and dynamics is crucial here to understand the physics of those complex systems under analysis. However, many times the information we can accede to about the actual topology of interactions is somehow incomplete, because of experimental limitations or because of lags on the details of the system \cite{yu06,nishikawa10,timme07,nitzan17}. Moreover, given that the only reflection of the dynamics on networks is usually a certain macroscopic observable, it turns out that many topologies are compatible with the same dynamical output, raising the problem of multi-valuation \cite{napoletani08,barzel09,eguiluz11} (i.e. different topologies with the same dynamical response). 


Following this perspective, we analyze the relation between function and structure in a novel mapping problem. Essentially, given a certain network structure and a dynamical process on top of it, we wonder how to transform the network into a different structural connectivity so that the collective behavior (i.e. the function) remains invariant. Such transformation must adjust the weights of the interactions in the new configuration to achieve the goal of having an equivalent steady-state functionality to the original structure. In this letter, we present a new formalism, based on the maximum entropy principle \cite{shannon49,jaynes57}, to derive analytical transformations for the resulting weights when only local information (at the nodes' scale) is available. Furthermore, we show that the mapping of homogeneous networks into heterogeneous ones is usually less accurate and requires more -costly- microscopic information than the reverse process, unveiling a \emph{symmetry-unbalance} phenomenon that emerges from the partial impossibility of preserving the local structural constraints. 

To derive the network transformations, we focused on a particular dynamical process, the synchronization of coupled phase oscillators. This paradigmatic example of emergent phenomena has been extensively studied \cite{acebron05,arenas08,rodrigues15}, to unveil fundamental aspects related to the mapping problem, such as the inference of structure from response dynamics \cite{arenas06a,arenas06b,timme07,prignano12}, the dependency of the collective behavior on the topology \cite{gardenes07a,rodrigues15,restrepo05} and the network optimization to maximize the stability of the fully-synchronized attractor \cite{motter05,zhou06b}. The Kuramoto model (KM) \cite{kuramoto03} consists of a population of $N$ coupled phase oscillators that evolve in time according to the set of equations
\begin{equation}
\dot{\theta}_i = \omega_i + K \sum_{j = 1}^N \lambda_{ij} \sin(\theta_j - \theta_i), \mbox{  } \forall i \mbox{ } \in N, 
\label{kuramoto} 
\end{equation} where $\theta_i$ is phase of the i-oscillator, $\omega_i$ its natural frequency, drawn from a probability distribution $g(\omega)$, $\lambda_{ij}$ are the elements of the coupling matrix $\mathbf{\Lambda}$ that capture the presence of a connection and its intensity and $K$ is a constant coupling strength that scales all the weights. The collective behavior of the KM is usually described through the complex order parameter 
\begin{equation}
re^{i\Psi(t)}= \frac{1}{N}\sum_{j = 1}^N e^{i\theta_{j}},
\label{order_parameter}
\end{equation} 
where the modulus $r$ measures the overall degree of synchrony and $\Psi(t)$ the average phase of the system. Here, we assume that the macroscopic order parameter $r$ is the only available observable from measurements, and we look for transformations of $\mathbf{\Lambda}$ that keep this observable invariant, for any value of the control parameter $K$. 

It is well known that particular unweighted instances drawn from the same degree distribution will produce the desired invariant collective behavior \cite{doro08,arenas08,newman10}. We wonder if the former invariance can be achieved for weighted networks drawn from different degree distributions, preserving the number of nodes $N$. We consider a target network $\mathcal{A}$ with a given coupling matrix $\mathbf{A}$, which might be non-symmetric and directed with fixed entries $\lambda_{ij}^A$, and a candidate network $\mathcal{B}$, with different coupling matrix $\mathbf{B}$ and $\lambda_{ij}^B$. We impose transformations of $\mathcal{B}$ in the form $\mathbf{B'}= \mathbf{W} \circ \mathbf{B}$, with entries $w_{ij} \lambda_{ij}^B$, where $w_{ij}$ are the parameters to find.  Note that we can absorb the weights of $\mathcal{B}$ into $\mathbf{W}$, keeping only the binary values $b_{ij}$ of the structure of $\mathcal{B}$. After this simplification, the entries of the transformed network can be written as $(\mathbf{B'})_{ij} =w_{ij} b_{ij}$. Furthermore, we assume that the $N$ units are distinguishable and preserve their intrinsic properties in the transformation ($\omega_i^A = \omega_i^{B'} \ \forall \mbox{ } i \in N$), which ensures that we are dealing with particular instances of networks and not with averaged ensembles. Then, the condition for functional synchronization invariance can be written as 
\begin{equation}
\langle r^2(\vec{\omega},K,\mathbf{A}) \rangle = \langle r^2(\vec{\omega},K,\mathbf{B'}) \rangle, \mbox{ } \forall \mbox{ } K > 0,
\label{invariance}
\end{equation} where the measurements are in the steady-state, the average refers to different initial conditions, accounting for fluctuations of order $1/\sqrt{N}$, and the parameters of the dynamical process ($\vec{\omega},K$) are fixed in both networks. 

Inspired by the derivation of statistical mechanics from information theory as a particular case of statistical inference, see \cite{jaynes57}, we tackle the functional mapping defined above as an optimization problem for the unknown weights subject to structural constraints on the networks that capture our prior knowledge on the system. The key assumption here is that Eq.(\ref{invariance}) can be achieved by imposing a local detailed balance for the main structural properties of the nodes: the overall coupling intensities received from neighbours (or input strengths \cite{newman10}). For each node, we define the zero-order input strength as $s_i^{(0)} = \sum_j \lambda_{ij}$, the first-order as $s_i^{(1)} = \sum_j \lambda_{ij} (\sum_k \lambda_{jk})$ and so on. For a fixed order $M$, the detailed balance is given by a set of $N(M+1)$ equations for the $s_i^{(m)}$. If we let $q$ be the N-vector of ones $q = (1,1,...,1)^\top$, we have
\begin{equation}
\mathbf{A}^{m+1} q = \mathbf{B'} \mathbf{A}^{m} q, \ \forall \ 0 \leq m \leq M 
\label{matrix}
\end{equation}
where $(\mathbf{A}^{m+1}q)_i = s_i^{(m)}$ are the node structural bounds in the optimization of the weights in $\mathbf{B'}$. The local constraint ($m = 0$) can be written explicitly as 
\begin{equation}
\sum_{j = 1}^N \lambda_{ij}^A= \sum_{j = 1}^N w_{ij}b_{ij}, \mbox{ } \forall \mbox{ } i \in N, 
\label{constraint_1}
\end{equation}
which ensures to preserve the overall coupling in the transformation ($\sum_i \sum_j \lambda_{ij}^A = \sum_i \sum_j w_{ij} b_{ij}$). The \emph{ansatz}  of Eq.(\ref{constraint_1}) relies on the weighted annealed approximation \cite{serrano05,baronchelli10}, that assumes statistical similarity among nodes with the same $s_i^{(0)}$. This description is known to be valid in the linear regimes of the diffusion of random walkers \cite{baronchelli10} and the Master Stability Function (MSF) \cite{motter05,zhou06b}. Here, if the coupling strength $K$ is sufficiently large, Eq.(\ref{kuramoto}) can be linearized, and using statistical and mean-field arguments \cite{zhou06b}, the system can be uncoupled, with each unit being driven only by its input strength. The underlying assumption is that higher order constraints ($m > 0$) might be required when the non-linearity of Eq.($\ref{kuramoto}$) plays a crucial role or the connectivity patterns are highly non-trivial (heterogeneity, correlations, etc...).

We take advantage of information theory \cite{shannon49}, to define an appropriate objective function to optimize the unknown weights. In an uncertainty scenario, the best we can do is to rely on the Maximum Entropy Principle \cite{jaynes57}. It states that, subject to the available data (i.e. the constraints in Eq.(\ref{matrix})), the probability distribution which best captures our lack of information is the one that maximizes the entropy. Here, we can interpret the weights distribution in probabilistic terms, where the input strength $s_i^{(0)}$ is the normalization condition, and we can define the entropy \cite{shannon49} of a node $S_i$ as a sum over the accessible states defined as those where $b_{ij} = 1$, 
\begin{equation}
S_i = - \sum_{j = 1}^Nw_{ij}\log w_{ij}, \mbox{ } \forall \mbox{ } i \in N, 
\label{entropy}
\end{equation}
where the normalization constant has been neglected for simplicity and it is assumed that $w_{ij} \geq 0$. We can use the method of Lagrange multipliers \cite{jaynes57} to solve this optimization problem. The lagragian function reads as 
\begin{eqnarray}
{\cal L} &= \sum_{i=1}^{N} (S_i - \sum_{m = 0}^M \beta_{i}^{(m)}[(\mathbf{A}^{m+1}q)_i - (\mathbf{B'} \mathbf{A}^{m} q)_i])\nonumber \\  
&
\label{lagrangian}
\end{eqnarray}
where $\beta_i^{(m)}$ is the $m$-order lagrange multiplier of $i$-node. By optimizing Eq.(\ref{lagrangian}) with respect to the unknown weights and finding the values of the multipliers, we can derive analytical expressions for the entries of $\mathbf{B'}$. For the zero-order case ($M = 0$), we obtain
\begin{equation}
w_{ij}^{(0)} = \frac{\sum_{k = 1}^N \lambda_{ik}^{A}}{\sum_{k = 1}^N b_{ik}}, \mbox{ } \forall \mbox{ } i,j \in N, 
\label{we_max}
\end{equation}
that can be written as $w_{ij}^{(0)} = s_i^{(0)}/k_i^B$, where $k_i^B$ is the degree of node $i$ in $\mathcal{B}$. This solution is very \emph{intuitive}, since it homogeneously allocates  the input strength of a node into the available links. The weights are therefore equal for all the incoming links of a node ($w_{ij}$ is independent of the node $j$), implying usually a non-symmetric coupling.

The solution in Eq.(\ref{we_max}) is precisely the scheme used in \cite{motter05,zhou06b} to transform a network topology into a purely homogenous one to optimize the stability of the synchronized state in the scope of the MSF. That means that the solution is valid in the linear regime, close to the synchronization attractor. However, this solution is yet to be validated in the fully non-linear regime. We simulate the dynamics of $N = 2000$ oscillators following Eq.(\ref{kuramoto}) with fixed $g(\omega) \in (-\pi,\pi)$, measuring $\langle r^2 \rangle$ in a quasi-static process controlled by the control parameter $K \in [0,0.5/N]$. We propose to map pairs of uncorrelated networks drawn from different degree distributions, that range from homogeneous in degree, Erd\"os-R\'enyi networks, to power-law in degree networks, which are initially unweighted and symmetric. We use the model in \cite{gardenes06a}, to interpolate between both degree distributions using a single parameter $\alpha$. For $\alpha =0$ we have pure power-law distributions $p(k)\sim k^{-\gamma}$ with exponent $\gamma=3$ while for $\alpha =1$ we obtain homogeneous random networks, keeping the average degree fixed, in our case $\langle k \rangle = 10$. The mapping transformation is then as follows: we fix the topologies of a network ${\mathcal A}$ drawn from the model for a certain value $\alpha$, i.e. the target network ${\mathcal A}_{\alpha}$, and the candidate network ${\mathcal B}_{\alpha'}$ drawn for another value $\alpha'$. Then, we compute the weights, using Eq.(\ref{we_max}), to map the candidate network into the target one and obtain the resulting $T_0({\mathcal B}_{\alpha'} | {\mathcal A}_{\alpha})$, where the subindex of $T$ refers to the fact that the method exploits only zero-order information. 

\begin{figure}[h]
\begin{center}
\includegraphics[width=1\columnwidth,angle=0]{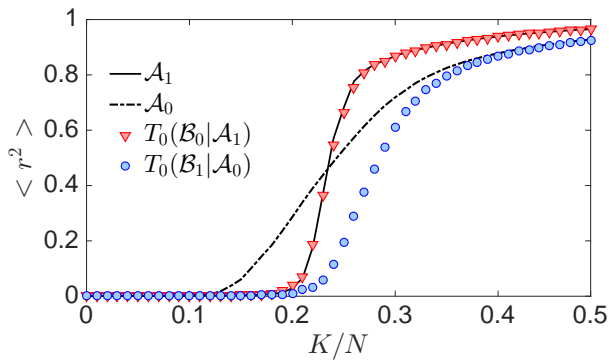}
\caption{
Synchronization diagram. We plot $r^2$ as a function of the coupling strength $K$ of the Kuramoto model, with $\Delta (K/N) = 0.01$ simulated with a 4th-order Runge--Kutta method with $\Delta t = 0.01$, for one instance of ${\mathcal A}_{1}$ (Erd\"os-R\'enyi) and ${\mathcal A}_{0}$ (power-law) networks and their respective transformations using Eq.(\ref{we_max}), averaged over 50 realizations with $\theta_0 \in [-\pi,\pi]$ (standard deviations are smaller than the size of the symbols).}
\label{fig:maxentropy}
\end{center}
\end{figure}

In Fig.(\ref{fig:maxentropy}) we present the results of the transformation for the extreme cases $T_0({\mathcal B}_{0}|{\mathcal A}_{1})$ and $T_0({\mathcal B}_{1}|{\mathcal A}_{0})$. The results evidence that the functional invariance is attained in the linear regime $(K \gg K_c)$ for both transformations. However, there is a clear discrepancy in the transformation $T_0({\mathcal B}_{1}|{\mathcal A}_{0})$, i.e. from a homogeneous in degree network towards an heterogeneous, power-law, network. This discrepancy shows that, when Eq.(\ref{we_max}) is applied, homogeneous networks are not able to capture the role of heterogeneous connectivity patterns.

\newpage
To improve the accuracy of the $T_0$ method in the mapping, we need to include higher-order constraints. We extend the detailed balance to a further order ($M = 1$) by imposing that, for each node, the transformation must also preserve the first-order input strengths $s_i^{(1)}$, i.e. 
\begin{equation} 
\sum_{j = 1}^N \lambda_{ij}^A s_j^{(0)} = \sum_{j = 1}^N w_{ij}b_{ij} s_j^{(0)}, \mbox{ } \forall \mbox{ } i \in N. 
\label{constraint_2}
\end{equation}
Note that $s_j^{(0)}$ is the same at both ends of Eq.(\ref{constraint_2}) because we still retain the constraint presented in Eq.(\ref{constraint_1}). We aim to maximize Eq.(\ref{entropy}) subject to Eq.(\ref{constraint_1}) and Eq.(\ref{constraint_2}). The lagrangian in Eq.(\ref{lagrangian}) can be written explicitly as 
\begin{eqnarray}
{\cal L} &= \sum_{i=1}^{N} [-\sum_{j=1}^{N} w_{ij}\log w_{ij} -\beta_i^{(0)} (s_i^{(0)} -\sum_{j=1}^{N} w_{ij}b_{ij}) \nonumber \\  
&- \beta_i^{(1)}( \sum_{j=1}^{N} \lambda_{ij}^{A} s_j^{(0)}-\sum_{j=1}^{N} w_{ij}b_{ij} s_j^{(0)})]. 
\label{lagrangian2}
\end{eqnarray}
By imposing $d{\cal L}/dw_{ij} = 0$ and isolating the unknown weight $w_{ij}$, we obtain the implicit expression 
\begin{equation}
w_{ij}^{(1)}(\beta_i) = \frac{s_i^{(0)}e^{-\beta_i s_j^{(0)}}}{\sum_{k = 1}^N b_{ik} e^{-\beta_i s_k^{(0)}}}, \mbox{ } \forall \mbox{ } i,j \in N.
\label{implicit}
\end{equation} 
The values of the multipliers $\beta_i$ are found by substituting Eq.(\ref{implicit}) back in Eq.(\ref{constraint_2}) and numerically solving the resulting system. However, the existence of real and non-negative solutions cannot be ensured a priori. Indeed, the structural bounds are easily estimated by considering the worst-case scenarios, i.e.
\begin{equation}
s_i^{(0)} \times \min_{\forall j \in N}(b_{ij} s_j^{(0)}) \leq s_i^{(1)} \leq s_i^{(0)} \times \max_{\forall j \in N} (b_{ij}s_j^{(0)}), \mbox{ } \forall \mbox{ } i \in N.
\label{ineq}
\end{equation}
The inequality in Eq.(\ref{ineq}) turns out to be unfeasible for most nodes if the reference network is very heterogeneous in local input strength. Let us illustrate this by considering, on one hand, that $\mathcal{A}$ follows a power-law distribution with $p(s) = c s^{-\gamma}$. Then, if network $\mathcal{B}$ is sufficiently well-connected $(k_i^B \gg 1 \mbox{ } \forall \mbox{ } i \in N)$ and assuming $N$ large, we can approximate the constraints by 
\begin{gather}
s_i^{(0)} \simeq k_i^B \int_{0}^{\infty} e^{-\beta_i s} p(s) ds = \frac{c k_i^B}{\beta_i^{1-\gamma}} \int_{0}^{\infty} e^{-x}x^{-\gamma} dx\\
s_i^{(1)} \simeq k_i^B \int_0^{\infty} s e^{-\beta_i s} p(s) ds = \frac{c k_i^B}{\beta_i^{2-\gamma}} \int_{0}^{\infty} e^{-x}x^{-\gamma+1} dx.
\end{gather} 
The first integral can be written as the Gamma function $\int e^{-x}x^{-\gamma} dx = \Gamma(1-\gamma)$. Using the well-known property $\Gamma(z+1) = z\Gamma(z)$ and dividing both equations, we obtain
\begin{equation}
\beta_i \simeq \frac{s_i^{(0)}}{s_i^{(1)}}(1-\gamma), \mbox{ } \forall \mbox{ } i \in N,
\end{equation}
which is negative for $\gamma = 3$, thus unveiling the structural restrictions that emerge when mapping any arbitrary network into a highly heterogeneous one. On the other hand, Eq.(\ref{we_max}) is recovered from Eq.(\ref{implicit}) only when $s_i^{(0)} \simeq \langle s^{(0)} \rangle, \mbox{ } \forall \mbox{ } i \in N$, i.e when $\mathcal{A}$ is very homogeneous in local input strength, regardless of the topology of $\mathcal{B}$.

The previous reasoning unfolds the \emph{symmetry-unbalance} observed in Fig.(\ref{fig:maxentropy}) and suggests that the mapping can indeed be enhanced, although it is strongly limited by the structural bounds. To provide an analytical transformation that improves the performance of Eq.(\ref{we_max}) while still preserving $w_{ij} \geq 0$, we expand Eq.(\ref{implicit}) to first order around its average value, i.e. 
\begin{equation}
w_{ij}^{(1)}(\beta_i) \simeq \frac{s_i^{(0)}[1-\beta_i(s_j - \langle s \rangle)]}{\sum_{k=1}^N b_{ik} [1 - \beta_i(s_k - \langle s \rangle)]}, \mbox{ } \forall \mbox{  } i, j \in N, 
\label{w_app0}
\end{equation}
where $\langle s \rangle = (1/k_i^B)\sum_j b_{ij}s_j^{(0)}$. We insert Eq.(\ref{w_app0}) into Eq.(\ref{constraint_2}) to obtain an approximate value $\beta_i^* \simeq \beta_i $ as 
\begin{equation}
\beta_i^* = \frac{1}{s_i^{(0)}}(\frac{s_i^{(0)}\langle s \rangle - s_i^{(1)}}{\langle s^2\rangle - \langle s \rangle^2}), \mbox{ } \forall \mbox{ } i \in N. 
\label{beta}
\end{equation}
The solution is finally obtained by direct substitution of Eq.(\ref{beta}) into Eq.(\ref{implicit}), and we denote this transformation  $T_1(\mathcal{B_{\alpha'}}|\mathcal{A_{\alpha}})$. Note that $T_1$ does not provide uniform weighting, but depends explicitly on the balance between input strengths and heterogeneity in each node. 

Now we can compare the performance of transformations $T_0$ and $T_1$ in the mapping. We define, for each transformation, the dynamical error \begin{equation}
\sigma_d = N^{-1} \int_{0}^{K_{\infty}} [\langle r^2(\vec{\omega},K,\mathbf{A}) \rangle -\langle r^2(\vec{\omega},K,\mathbf{B'}) \rangle ]^2dK, 
\end{equation}
as a measure of the total difference in the synchronization diagrams between the target and transformed networks, and we define the structural error 
\begin{equation}
\sigma_s = N^{-1} \sum_{i}^N [\sum_{j}^N (\lambda_{ij}^As_j^{(0)} - w_{ij}b_{ij}s_j^{(0)})]^2,
\end{equation}
as a measure of the total difference in the first-order local structure. 
\begin{figure}[h]
\begin{center}
(a)
\includegraphics[width=1\columnwidth,angle=0]{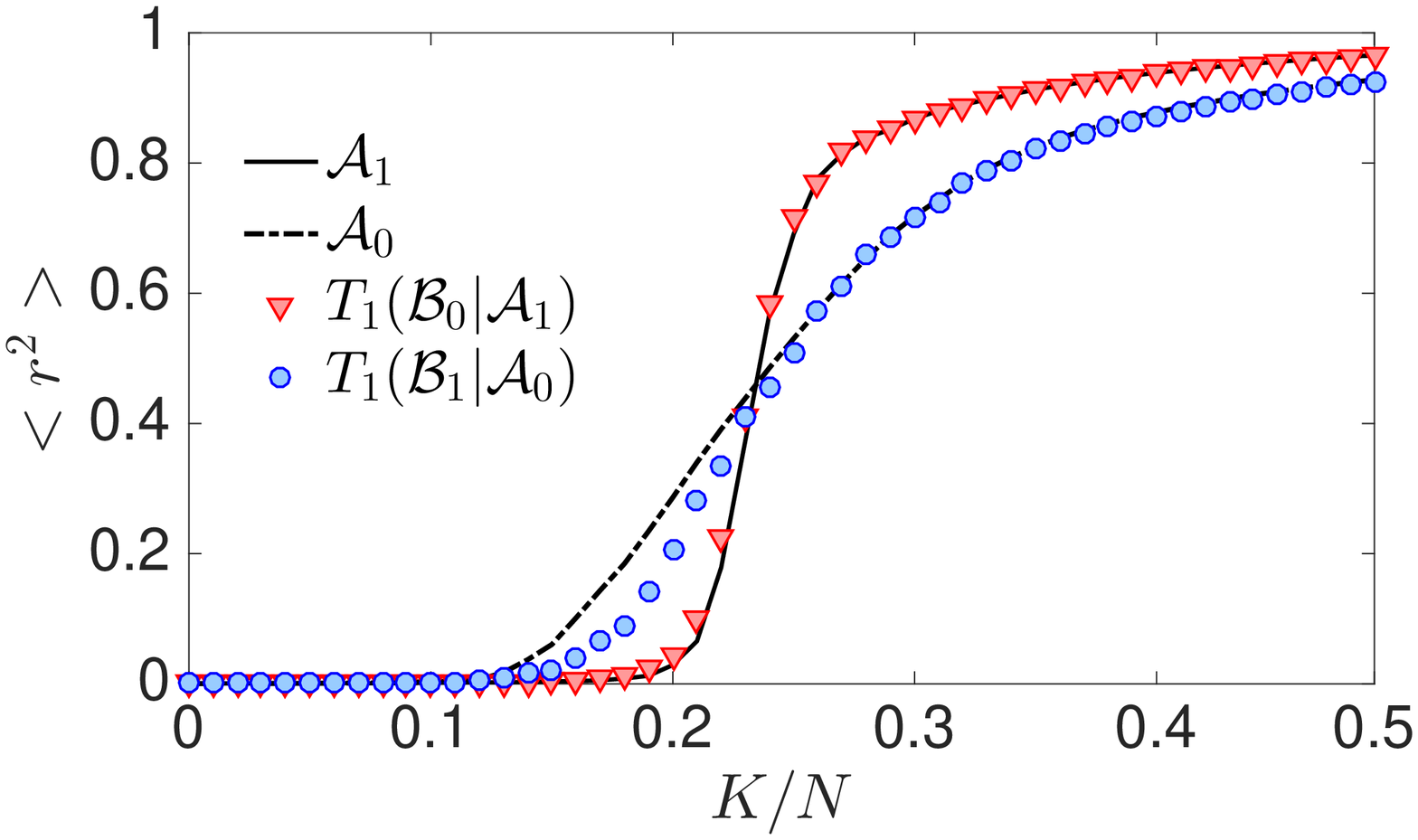}
(b)
\includegraphics[width=0.46\columnwidth,angle=0]{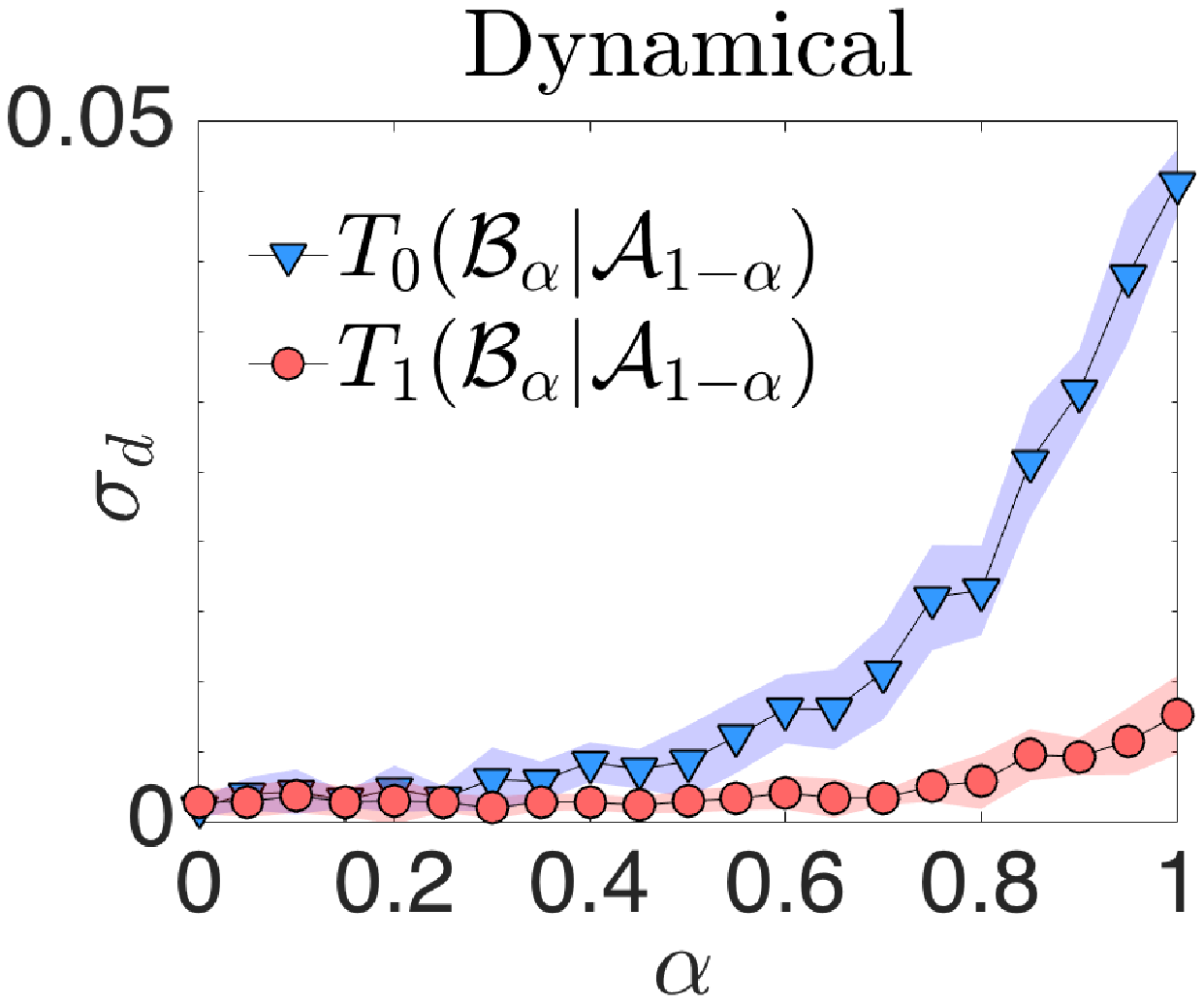}
\includegraphics[width=0.46\columnwidth,angle=0]{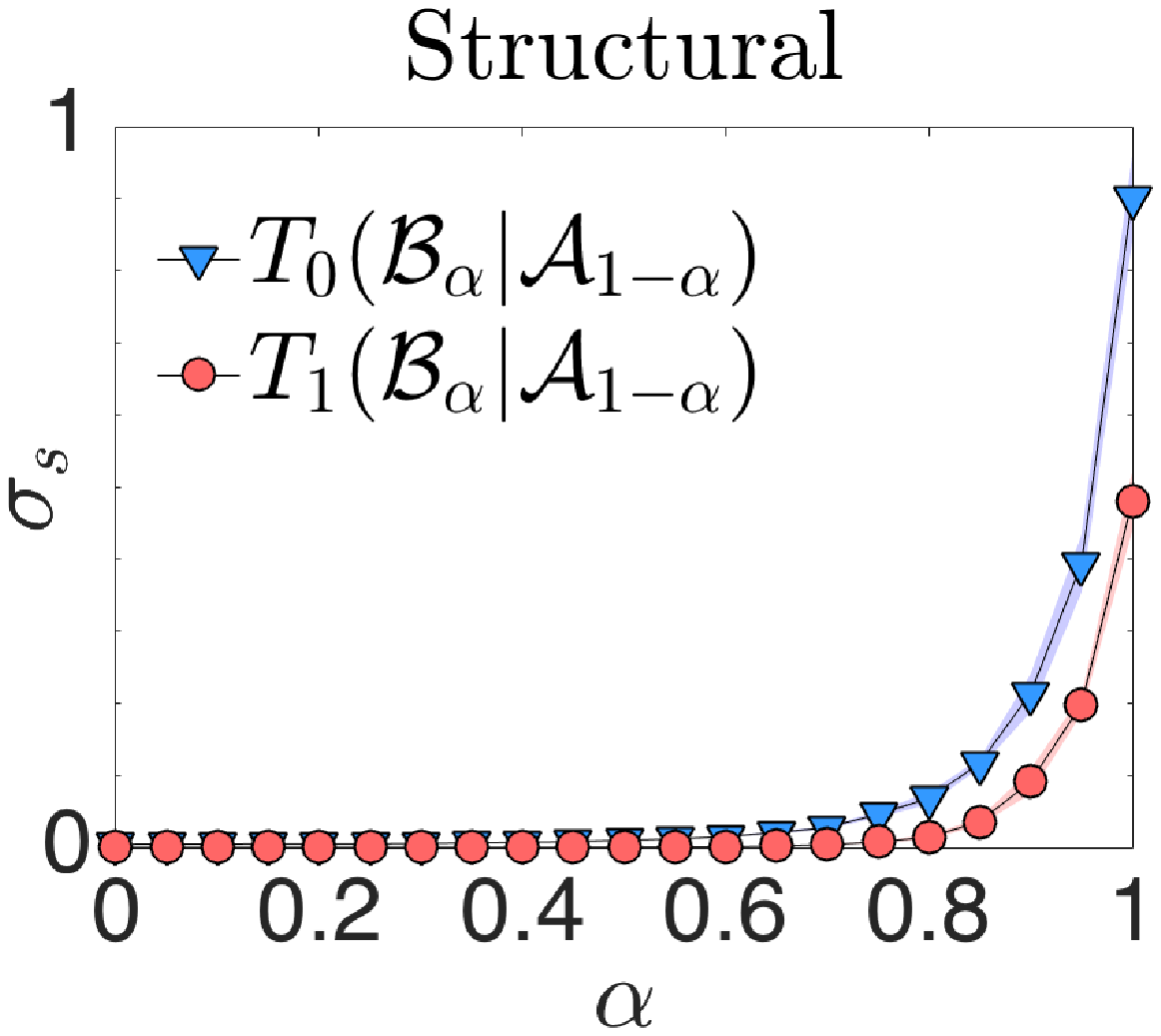}
\caption{a) Synchronization diagram. We plot $r^2$ as a function of $K$, for one instance of ${\mathcal A}_{1}$ and ${\mathcal A}_{0}$ networks and transformations $T_1(\mathcal{B}_0|\mathcal{A}_1)$ and $T_1(\mathcal{B}_1|\mathcal{A}_0)$ using Eqs.(\ref{implicit},\ref{beta}), averaged over 50 realizations with $\theta_0 \in [-\pi,\pi]$. b) Dynamical (left) error curves for $T_0(\mathcal{B}_\alpha |\mathcal{A}_{1-\alpha})$ and $T_1(\mathcal{B}_\alpha |\mathcal{A}_{1-\alpha})$, averaged over 100 independent network instances for each $\alpha$ (standard deviations fall in the shaded region). In b) right, associated structural error curves (standard deviations are of the size of the symbols and the values of $\sigma_s$ are properly normalized).
}
\label{fig:maxentropy2}
\end{center}
\end{figure}
In Fig.(\ref{fig:maxentropy2}a) we present the synchronization diagram for the extreme cases $T_1({\mathcal B}_{0}|{\mathcal A}_{1})$ and $T_1({\mathcal B}_{1}|{\mathcal A}_{0})$ in the same set up as before ($N = 2000$). We can observe a significant improvement in the transformation $T_1({\mathcal B}_{1}|{\mathcal A}_{0})$ with respect to the zero-order method in Fig.(\ref{fig:maxentropy}), although there still are non-vanishing errors around the critical point due to the unfeasible structural bounds of Eq.(\ref{ineq}). In Fig.(\ref{fig:maxentropy2}b), we plot the dynamical $\sigma_d$ and structural $\sigma_s$ errors for different values of the parameter $\alpha$ in $T(\mathcal{B}_\alpha|\mathcal{A}_{1-\alpha})$. Note how the accuracy of the transformations is enhanced by $T_1$ for any value of $\alpha$, and it is associated to a decrease in the structural error, thus validating the main assumptions of our approach. 

Furthermore, the approximate solution of Eqs.(\ref{implicit},\ref{beta}) can still be improved by i) considering higher-order constraints $(M > 1)$, but then the system would become coupled and it should be solved simultaneously for all nodes, ii) extending the expansion of Eq.(\ref{implicit}) with additional terms, iii) allowing the presence of negative interactions or indistinguishable units (without labelling the nodes in the transformation), and also iv) imposing global constraints instead of local ones (requiring costly numerical methods and global objective functions \cite{dedomenico16b}). 

Summarizing, we have presented an analytical methodology that successfully produces functional synchronization invariant networks for the KM, by transforming the weights of the interactions, while preserving the underlying topologies, and exploiting only local structural information. We have shown that different microscopic configurations can produce the same macroscopic dynamical observables if the weights are adjusted in a way that the main local properties of the nodes are preserved. Furthermore, we have unveiled that the mapping of homogeneous networks into heterogeneous ones requires to exploit additional (up to first-order) information and it is more complicated than the reverse process, due to intrinsic structural limitations of the networks. 

The presented formalism can be applied in a wide spectra of problems beyond the mapping scenario. Our framework provides a more comprehensive understanding of the collective behavior of oscillators on weighted and directed networks from a local perspective and can be used to make analytical predictions on them (when transformed to unweighted structures) \cite{rodrigues15,restrepo05}. Also, the transformations can induce specific features of heterogeneous networks in homogeneous ones and vice-versa, without changing the underlying structure. Straightforward examples include the possibility to induce explosive transitions in homogeneous networks (by correlating the intrinsic frequencies with the input strengths \cite{gardenes11a}) and to control the critical point of a macroscopic phase transition \cite{doro08,rodrigues15} only by a local readjustment of weights. From a theoretical point of view, our results are sheltered by previous works that explore information-theoretic tools to study the structure of complex networks \cite{park04,anand09,anand11} and to tackle reconstruction problems \cite{rosvall07,squartini11,mastrandrea14}. Nevertheless, here we introduce a novel connection between purely structural constraints and collective dynamical behavior. This new connection can help in refining state-of-art inference methods with driving signals \cite{timme07,nitzan17} (by inferring appropriate network candidates from the available structural and dynamical information), it deeps our understanding on findings that relate weighted, directed and inhibitory interactions to optimal synchronization performance \cite{korniss07,belykh08,skardal14}, and provides a new approach for evolving networks models \cite{doro08,newman10,rodrigues15}, in which a network of biological units might evolve, due to an evolutionary pressure, towards heterogeneous structures that maximize the number of accesible transformations and, consequently, their potential dynamical range \cite{larremore11}. 

\acknowledgments{L. A.-F. thanks G. Mosquera-Do\~nate for proposing the method of Lagrange Multipliers and B. Steinegger and A. Arola for fruitful discussions. L. A.-F. and A. A. acknowledges the Spanish MINECO, Grant No. FIS2015-71582-C2-1. A. A acknowledges funding also from ICREA Academia and the James S. McDonnell Foundation. A. D.-G. acknowledges financial support from MINECO (FIS2015-71582-C2-2), and the Generalitat de Catalunya (Project No. 2014SGR-608).}


\end{document}